\documentclass[a4paper,12pt]{article}
\title{Identifying Clusters in Bayesian Disease Mapping}
\author{Craig Anderson, Duncan Lee \& Nema Dean}
\date{}

\newcommand{\bd}[1]{\ensuremath{\mbox{\boldmath $#1$}}}
\usepackage{natbib}
\usepackage{graphicx}

\begin{document}

\maketitle

\footnotetext{To whom correspondence should be addressed.}

\begin{abstract}
{Disease mapping is the field of spatial epidemiology interested in estimating the spatial pattern in disease risk across $n$ areal units. One aim is to identify units exhibiting elevated disease risks, so that public health interventions can be made. Bayesian hierarchical models with a spatially smooth conditional autoregressive prior are used for this purpose, but they  cannot identify the spatial extent of high-risk clusters. Therefore we propose a two stage solution to this problem, with the first stage being a spatially adjusted hierarchical agglomerative clustering algorithm. This algorithm is applied to data prior to the study period, and produces $n$  potential cluster structures for the disease data.  The second stage fits a separate Poisson log-linear model to the study data for each cluster structure, which allows for step-changes in risk where two clusters meet. The most appropriate cluster structure is chosen by model comparison techniques, specifically by minimising the Deviance Information Criterion. The efficacy of the methodology is established by a simulation study, and is illustrated by a study of respiratory disease risk in Glasgow, Scotland. }
{Clustering, Conditional autoregressive model, Disease mapping.}
\end{abstract}

\section{Introduction}
\label{sec:Introduction}
In epidemiology it is well known that the risk of disease varies across space, due to differences in environmental exposures and the risk inducing behaviour of the population. One of the main causes of the latter is poverty, with a recent Audit Scotland report (\cite{audit12}) finding that the average life expectancy for men is 10.9 years less (70.1 compared 81.0) for the most deprived communities compared with the most affluent. The extent and pattern of such health inequalities are illustrated via disease maps, which are produced by first partitioning the study region into $n$ non-overlapping areal units such as electoral wards or census tracts. The overall risk of disease is then computed and mapped for the population living in each areal unit, and health agencies have produced such maps for numerous diseases including cancer (\cite{phe10}) and cardiovascular disease (\cite{cdc11}). The key benefit of such maps is that they allow public health officials to identify  clusters of areal units that exhibit elevated disease risks, which in turn enables interventions to be appropriately targeted at the communities at greatest need. Such interventions can include a vaccination programme, or a public awareness campaign about potential risk factors. Thus in  addition to the obvious public health benefit, the identification of high-risk clusters through the use of disease maps can help to reduce health care costs.\\

A number of different approaches have been proposed for mapping the spatial pattern in disease risk and identifying high-risk clusters, including Bayesian hierarchical modelling (\cite{charras12}), scan statistics (\cite{kulldorff97}) and point process methodology (\cite{diggle05}). The first of these is typically based on a Poisson log-linear model, where the spatial pattern in disease risk is represented by covariates and/or a set of random effects.  The latter are included to account for spatial correlation in the response not captured by the covariates, and are typically modelled by a conditional autoregressive (CAR) prior.  These priors were proposed by \cite{besag91} and developed by  \cite{leroux99}, and are a type of Gaussian Markov random field. These priors naively assume that all pairs of random effects in geographically adjacent areal units are correlated, thus producing a spatially smooth map of disease risk. While such spatial smoothing facilitates the borrowing of strength in the estimation of disease risk between neighbouring areas, it is contrary to the goal of identifying a high risk cluster, which exhibits a markedly increased risk of disease compared with its surrounding areas.\\

Therefore, the contribution of this paper is the development of new methodology to estimate the spatial pattern in disease risk and detect the spatial extent of high and low risk clusters. The methodology is a fusion of hierarchical agglomerative clustering techniques with conditional autoregressive models, and takes the form of a two-stage approach. The first stage is a hierarchical agglomerative clustering algorithm, that is extended to respect the spatial contiguity structure of the study region. This algorithm is applied to disease data preceding the study period, which elicits $n$ prior candidate cluster configurations containing between 1 and $n$ clusters.  The second stage is to fit a separate Bayesian Poisson log-linear model to the study data for each candidate cluster structure, which models disease risk as spatially smooth within a cluster with a potential step-change between clusters. The final cluster structure  is chosen by minimising the Deviance Information Criterion (DIC, \cite{spiegel02}), a model comparison statistic.\\

The hierarchical agglomerative clustering algorithm is not applied to the study data itself, because this would use the information in the data twice, once for eliciting a set of candidate cluster configurations and again for estimating the model parameters. Thus a second data set is required for the clustering stage, and possibilities include data on disease risk prior to the study period or data on a different disease. We utilise the former in this study, because the spatial patterns in the population characteristics governing disease risk (such as poverty) are unlikely to have changed greatly from year to year unless substantial urban regeneration has taken place.  We note that our approach is appropriate for data on chronic diseases whose risk factors are spatially stable, while it is unsuitable for epidemic diseases such as influenza, where the spatial pattern in disease risk before and during the epidemic would likely be different. The remainder of this paper is organised as follows.  Section 2 gives a brief introduction to Bayesian disease mapping, and critiques existing approaches to cluster detection in this context. Section 3 proposes our methodological development, while Section 4 establishes its efficacy via simulation. Section 5 presents the application that motivated our methodology, which is a study of respiratory ill health in Greater Glasgow, Scotland in 2011. Finally,  the implications of this paper and future work are discussed in Section 6.

\section{Bayesian disease mapping}
\label{sec:Background}

\subsection{Study Design and Modelling}
\label{sec:Design}

The study region $\mathcal{A}$ is partitioned into $n$ non-overlapping areal units $\mathcal{A}$ = \{$\mathcal{A}_1,\ldots,\mathcal{A}_n$\}, and $\bd{Y} = (Y_1,\ldots,Y_n)$ and $\bd{E} = (E_1,\ldots,E_n)$  represent the observed and expected numbers of disease cases in each unit during the study period. The latter are constructed by external standardisation, based on the age and sex demographics of the population living in each areal unit.   A Poisson log-linear model is commonly used to estimate disease risk, and a general form is given by

\begin{eqnarray}
\label{eq:PoissonLL}
Y_i | E_i, R_i &\sim& \mbox{Poisson}(E_{i}R_{i}) \hspace{10mm} i=1,...,n ,\\
\ln(R_i) &=& \mathbf{x}_i^T\bd{\beta} + \phi_i. \hspace{10mm}\nonumber
\end{eqnarray}

Here $R_i$ represents disease risk in areal unit $\mathcal{A}_i$, and is modelled by a vector of covariates $\mathbf{x}_i^T = (1, x_{i1},\ldots,x_{ip})$, with coefficients $\bd{\beta} =(\beta_0,\ldots,\beta_p)$, and a random effect $\phi_i$.  The random effects $\bd{\phi}=(\phi_i, \ldots, \phi_n)$ account for the unexplained spatial autocorrelation induced into the disease data by numerous factors, including unmeasured confounding, neighbourhood effects and grouping effects.  They are modelled by a conditional autoregressive (CAR) prior, which induces spatial correlation via a binary neighbourhood matrix $W$, where $w_{ij}=1$ if areal units $(\mathcal{A}_{i}, \mathcal{A}_{j})$ share a common border (denoted $i\sim j$) and $w_{ij}=0$ otherwise. Note that $w_{ii} = 0$ for all \emph{i}. CAR priors can be specified as a set of $n$ univariate conditional distributions $f(\phi_i|\bd{\phi}_{-i})$, where $\bd{\phi}_{-i} = (\phi_1,\ldots,\phi_{i-1},\phi_{i+1},\ldots,\phi_n)$. The simplest CAR prior is the intrinsic model  proposed by \cite{besag91}, and is given by

\begin{equation}
\label{eq:ICAR}
\phi_i|\bd{\phi}_{-i} \sim \mbox{N}\left(\frac{\sum_{j=1}^n w_{ij}\phi_j}{\sum_{j=1}^n w_{ij}},\frac{1}{\tau(\sum_{j=1}^n w_{ij})}\right) \mbox{ } i=1,\ldots,n,
\end{equation}

\noindent where $\tau$ is a conditional precision parameter.  The conditional expectation of $\phi_i$ is the mean of the random effects in neighbouring areal units, while the variance is inversely proportional to the number of neighbouring units.  This set of conditional distributions correspond to a multivariate Gaussian distribution, with mean zero but an improper precision matrix given by $Q=\tau(\mbox{diag}(W\mathbf{1})-W)$, where $W\mathbf{1}$ is a vector containing the number of neighbours for each areal unit.

\subsection{Literature review}
\label{sec:LitRev}
Research on the identification of step-changes in disease risk between geographically adjacent areal units has followed two main themes. The first has identified discontinuities in disease risk between geographically adjacent areal units, and a number of approaches based on the Bayesian hierarchical model outlined above have been proposed. The majority treat the elements of the neighbourhood matrix \{$w_{ij}|i \sim j$\} as binary random quantities, where estimating $w_{ij}$ = 0 corresponds to the identification of a boundary between $(\mathcal{A}_{i}, \mathcal{A}_{j})$ because ($\phi_i, \phi_j$) are conditionally independent and are not smoothed over in the modelling process. One of the first examples in this vein was \cite{lu07}, who proposed a logistic regression model for  \{$w_{ij}|i \sim j$\} using a measure of dissimilarity between $(\mathcal{A}_{i}, \mathcal{A}_{j})$ as the covariate. However, this results in an excessively large number of parameters, and to rectify this \cite{leemitch12} treated \{$w_{ij}|i \sim j$\} as a deterministic function of a small number of parameters and the areal level measure of dissimilarity. The same authors (\cite{lee2013}) also proposed an approach that iteratively re-estimates $\{w_{ij}|i\sim j\}$ and the remaining model parameters conditional on the other until a convergence criterion is reached. Finally, \cite{li11} fitted models with different $W$ specifications, hence different potential sets of  boundaries, and chose the best model by minimising the Bayesian Information Criterion (BIC).\\

The above  literature produces open boundaries, which are a set of potentially disjoint boundary segments that do not necessarily completely enclose an areal unit or group of units. In contrast, the field of cluster detection identifies  groups of areal units that exhibit substantially different risks compared to their neighbours, and thus the boundary surrounding them is closed. One of the first and still most widely used cluster detection approaches are scan statistics (\cite{kulldorff97}), which identify clusters of areal units that exhibit an elevated risk of disease. Their popularity in part stems from the availability of the SaTScan software, which allows others to easily implement this approach. However, scan statistics merely identify high-risk clusters, and are not suitable for estimating the spatial pattern in disease risk. This has led to a number of hierarchical modelling approaches such as \cite{knorr00}, \cite{green02}, \cite{charras12} and \cite{wakefield13} being proposed, and a first comparison to scan statistics is given by \cite{charras13}. This paper, together with \cite{charras12}, also assesses the utility of identifying clusters by applying a post processing clustering algorithm to a continuous disease map, although the spatial contiguity of the clusters is not guaranteed.\\

One of the main differences between these approaches is that \cite{knorr00} and \cite{wakefield13} force the clusters to be spatially contiguous, while \cite{green02} and \cite{charras12} do not. However, in all cases disease risk is assumed to be constant within a cluster, which has the advantage that it partitions the relative risk into risk classes/clusters which are easy to interpret for epidemiologists. However, for real data it is likely that disease risk varies within a cluster, and the model proposed here allows for such within cluster variation.  The other disadvantage of the above approaches is that they involve computationally complex estimation approaches, such as reversible jump Markov chain Monte Carlo algorithms (e.g.  \cite{knorr00}) or the Monte Carlo Expectation-Maximisation algorithm (e.g. \cite{charras12}). Such approaches are beyond the scope of most epidemiologists, and no publicly available software exists to allow others to implement them. The complexity of these estimation procedures is necessitated by them  estimating disease risk and the number of clusters simultaneously, where as here we propose a conceptually simpler two-stage approach utilising a model comparison approach. Both approaches have their own merits, as while our approach is simpler to implement for others (we provide R code to implement our approach in the supplementary material available at Biostatistics online), it does ignore the uncertainty about the number of clusters in the estimation procedure.

\section{Method}
\label{sec:Method}
We propose a two-stage approach for estimating the spatial pattern in disease risk and identifying spatially contiguous clusters that exhibit either elevated or reduced disease risks.   In the first stage (Section \ref{sec:Cluster}) we propose a spatially adjusted hierarchical agglomerative clustering algorithm, and use it to elicit a set of candidate cluster configurations for the data. In the second stage (Section \ref{Sec:Model}) we fit a separate Bayesian hierarchical model to  the study data for each cluster configuration, and choose the final cluster structure using the DIC.

\subsection{Stage 1 - Eliciting cluster configurations using hierarchical agglomerative clustering}
\label{sec:Cluster}
The method of clustering (for details see \cite{hastie01}) groups together objects that are similar whilst separating those that are different, which  is appropriate here because we wish to identify groups of areal units with similar disease risks.  We apply the clustering algorithm to disease data preceding the study period, because it is likely to exhibit a similar spatial risk pattern to the study data unless substantial urban regeneration has taken place. Let $(\bd{Y}^{(1)},\bd{E}^{(1)}),\ldots,(\bd{Y}^{(q)},\bd{E}^{(q)})$ denote the observed and expected disease counts for the $q$ time intervals (usually years) preceding the study period. We use these earlier data to elicit a set of $n$ potential cluster configurations for the study data, which are denoted here by $\{\mathcal{C}_1,\ldots,\mathcal{C}_n\}$.  Here $\mathcal{C}_k = \{\mathcal{C}_k(1),\ldots, \mathcal{C}_k(k)\}$ partitions the $n$ areal units $\mathcal{A}=\{\mathcal{A}_1, \ldots, \mathcal{A}_n\}$ into $k$ spatially contiguous groups, where $\mathcal{C}_k(j)$ is the \emph{j}th cluster. The motivation for this step is that the set of all possible spatially contiguous cluster configurations for the study region $\mathcal{A}$ is very large, so we use these earlier data to vastly reduce the number of potential cluster structures to be compared in stage 2.\\

The data are clustered on the log standardised incidence ratio scale, that is $\mbox{ln}(\bd{Y}^{(j)}/\bd{E}^{(j)})$, because it corresponds to the linear predictor scale in (\ref{eq:PoissonLL}).  Let $\bd{\psi}=[\mbox{ln}(\bd{Y}^{(1)}/\bd{E}^{(1)}),\ldots,\mbox{ln}(\bd{Y}^{(q)}/\bd{E}^{(q)})]$ be the ${n\times q}$ matrix whose columns comprise $\mbox{ln}(\bd{Y}^{(j)}/\bd{E}^{(j)})$ for $j=1,\ldots,q$, and denote the $i$th row by $\bd{\psi}_{i}=[\ln(Y_{i}^{(1)}/E_{i}^{(1)}),\ldots,\ln(Y_{i}^{(q)}/E_{i}^{(q)})]$, the vector of $q$ values for areal unit $\mathcal{A}_{i}$. We cluster the data using a modified hierarchical agglomerative clustering algorithm,  which initially considers each data point as its own singleton cluster, and then joins together the two least dissimilar clusters at each stage to form a larger cluster. This process is repeated until only one cluster containing all data points remains. For a configuration with $k$ clusters the dissimilarity, $d_{ij}$, between clusters \emph{i} $\left(\mathcal{C}_k(i)\right)$  and  \emph{j} $\left(\mathcal{C}_k(j)\right)$ can be measured by a number of metrics called linkage methods, and in this paper we consider the following three.
\begin{itemize}

\item Single linkage measures the dissimilarity as the shortest distance between two clusters, that is $d_{ij} = \mbox{min}\{||\bd{\psi}_{f}- \bd{\psi}_{g}||:  \mathcal{A}_f \in \mathcal{C}_k(i), \mathcal{A}_g \in \mathcal{C}_k(j)\}$, where $||.||$ denotes Euclidean distance.

\item Centroid linkage measures the dissimilarity as the Euclidean distance between the average of the two clusters, that is $d_{ij} = || \bar{\mathcal{C}}_k(i)- \bar{\mathcal{C}}_k(j)||$, where $\bar{\mathcal{C}}_k(i) = (1/n_i)\sum\limits_{f:\mathcal{A}_f \in \mathcal{C}_k(i)} \bd{\psi}_{f}$, and $n_i$ is the number of areal units in cluster $\mathcal{C}_k(i)$.

\item Ward's Linkage measures the dissimilarity as the increase in the error sum of squares (ESS) when joining two smaller clusters into a larger cluster, that is $d_{ij} = \mbox{ESS}(\mathcal{C}_k(i,j)) - \left[\mbox{ESS}(\mathcal{C}_k(i)) + \mbox{ESS}(\mathcal{C}_k(j))\right]$, where $\mathcal{C}_k(i,j) = \mathcal{C}_k(i) \cup \mathcal{C}_k(j)$ and $\mbox{ESS}(\mathcal{C}_k(i)) = \sum\limits_{f:\mathcal{A}_f \in \mathcal{C}_k(i)} ||\bd{\psi}_{f}  - \bar{\mathcal{C}}_k(i)||^2$.
\end{itemize}

We extend the hierarchical agglomerative clustering algorithm described above so that it produces spatially contiguous clusters, which is achieved by only allowing clusters that share a common border to be merged at each step.  The algorithm produces a set of candidate cluster structures $\{\mathcal{C}_1,\ldots,\mathcal{C}_n\}$ as follows:\\

\underline{Algorithm}

\begin{enumerate}
\item Construct $\mathcal{C}_n = \{\mathcal{C}_n(1),\ldots,\mathcal{C}_n(n)\}$, an initial cluster structure where each areal unit is in its own singleton cluster.

\item Repeat the following steps for $h=n,\ldots,2$, where step $h$ produces $\mathcal{C}_{h-1}$ from $\mathcal{C}_{h}$.

\begin{enumerate}
\item \begin{flushleft} Compute the $h \times h$ distance matrix D, whose\emph{ kl}th element is given by\linebreak
$$D_{kl} = \left\{
  \begin{array}{l l}
  |d_{kl}| &\quad \mbox{if } k \sim l \, \& \, k>l\\
  \infty & \quad \mbox{otherwise,}\\
  \end{array} \right.
$$
\linebreak where $d_{kl}$ is the distance between clusters ($\mathcal{C}_h(k),\mathcal{C}_h(l)$) under the selected linkage method, and $k \sim l$ means that the clusters contain at least one pair of areas that share a common border.

\end{flushleft}
\item Set $\{i,j\} = \arg \min (D_{kl})$, that is the identifiers of the two clusters that have the minimum dissimilarity as measured by the linkage method. In case of ties, $\{i,j\}$ is randomly selected from these.
\item
\begin{flushleft}Compute \footnotesize $$\mathcal{C}_{h-1} = \{\mathcal{C}_{h}(1), \ldots, \mathcal{C}_{h}(i-1), \mathcal{C}_{h-1}(i), \mathcal{C}_{h}(i+1), \ldots, \linebreak \mathcal{C}_{h}(j-1), \mathcal{C}_{h}(j+1), \ldots, \mathcal{C}_{h}(h) \},$$

\noindent\normalsize where $\mathcal{C}_{h-1}(i) = \mathcal{C}_{h}(i) \cup \mathcal{C}_{h}(j)$.
\end{flushleft}
\end{enumerate}
\end{enumerate}

\subsection{Stage 2 - Estimating the cluster structure using model comparison techniques}
\label{Sec:Model}
The study data are denoted by $(\bd{Y}, \bd{E})$, and the best cluster structure for these data from the set of $n$ candidates $\{\mathcal{C}_1, \ldots, \mathcal{C}_n\}$ elicited from stage 1 is estimated using a model comparison procedure. Specifically, the Bayesian Poisson log-linear model described below is fitted to the data based on each cluster configuration $\mathcal{C}_{k}$ in turn, and the cluster structure in the data is estimated by choosing the model that minimises the DIC. The DIC is defined as DIC = $\bar{D} + p_d$, where $\bar{D}$ is the mean posterior deviance and $p_d$ is the effective number of parameters, the latter acting as a penalty to penalise models with an excessively large number of parameters. For a given cluster structure $\mathcal{C}_{k}$, the proposed model is given by

\begin{eqnarray}
\label{eq:ClustModel}
Y_i | E_i, R_i &\sim& \mbox{Poisson}(E_{i}R_{i}) \hspace{10mm} i=1,\ldots,n, \nonumber\\
\ln(R_i) &=& \phi_i + \sum_{j=1}^k I[\mathcal{A}_i \in \mathcal{C}_k(j)]\alpha_j, \nonumber \\
\alpha_j &\sim& \mbox{N}(0,10) \hspace{10mm} j=1,\ldots,k,\\
\phi_i|\bd{\phi}_{-i} &\sim& \mbox{N}\left(\frac{\sum_{j=1}^n w_{ij}\phi_j}{\sum_{j=1}^n w_{ij}},\frac{1}{\tau (\sum_{j=1}^n w_{ij})}\right),\nonumber\\
\tau &\sim& \mbox{Gamma}(1,1).\nonumber\\
\nonumber
\end{eqnarray}

This model allows disease risk to evolve smoothly within a cluster with a disjoint multiplicative jump between clusters, which is achieved by combining the smooth intrinsic CAR model (\ref{eq:ICAR}) for $\bd{\phi}$ with a piecewise constant cluster model defined by  $\sum_{j=1}^k I[\mathcal{A}_i \in \mathcal{C}_k(j)]\alpha_j$ on the linear predictor scale.  Here, $I[.]$ denotes an indicator function, so that $I[\mathcal{A}_i \in \mathcal{C}_k(j)]$ equals one if areal unit $\mathcal{A}_i$ lies in cluster $j$ and is zero otherwise. Thus this piecewise constant cluster model is a single categorical covariate with $k$ levels, where each cluster represents a different level. We note that when areal unit $\mathcal{A}_{i}$ is in a singleton cluster, then this model essentially includes an indicator variable for that areal unit, resulting in the fitted value equalling the observed value. We considered modelling the cluster parameters $(\alpha_{1},\ldots,\alpha_{k})$ as random rather than fixed effects, but an initial simulation study showed that this resulted in poor performance in terms of cluster identification. Finally, the hyperparamters $(1,1)$ in the gamma prior will be varied in the simulation study, to gauge the sensitivity of the results.\\

Inference for the above model is implemented using integrated nested Laplace approximations (INLA, \cite{rue09}), because fitting the $n$ models corresponding to $\{\mathcal{C}_1, \ldots, \mathcal{C}_n\}$ would be computationally prohibitive using Markov Chain Monte Carlo (MCMC) methods. Inference using INLA has been shown by \cite{schrodle11} to produce almost identical results to MCMC simulation. The model above does not include additional covariates other than the factor variable representing the cluster structure, because the goal of the analysis is to identify clusters in the disease risk surface, not in the residual surface after adjusting for covariate factors.\\

\section{Simulation study}
\label{sec:Simulation}
A simulation study was conducted to establish the efficacy of the two-stage modelling approach outlined in the previous section.   The template for the study was the set of 271 Intermediate Geographies comprising the Greater Glasgow and Clyde health board, which is the study region for the motivating application presented in Section 5.  First, a preliminary study comparing the relative performances of the three linkage methods described in Section 3 was undertaken, and the results are summarised in Section 2 of the supplementary material available at Biostatistics online. The results show that centroid linkage always outperforms single and Ward's linkage methods, and is thus used throughout the remainder of this paper. A second study was then conducted comparing the two-stage approach proposed here with existing alternatives, and the results are summarised below. Finally, a number of sensitivity analyses were also conducted, which are summarised in Sections 3 to 5 of the supplementary material available at Biostatistics online.

\subsection{Data Generation}
\label{sec:ClustDataGen}
\begin{figure}[h!]
\begin{center}
\includegraphics[scale=0.45]{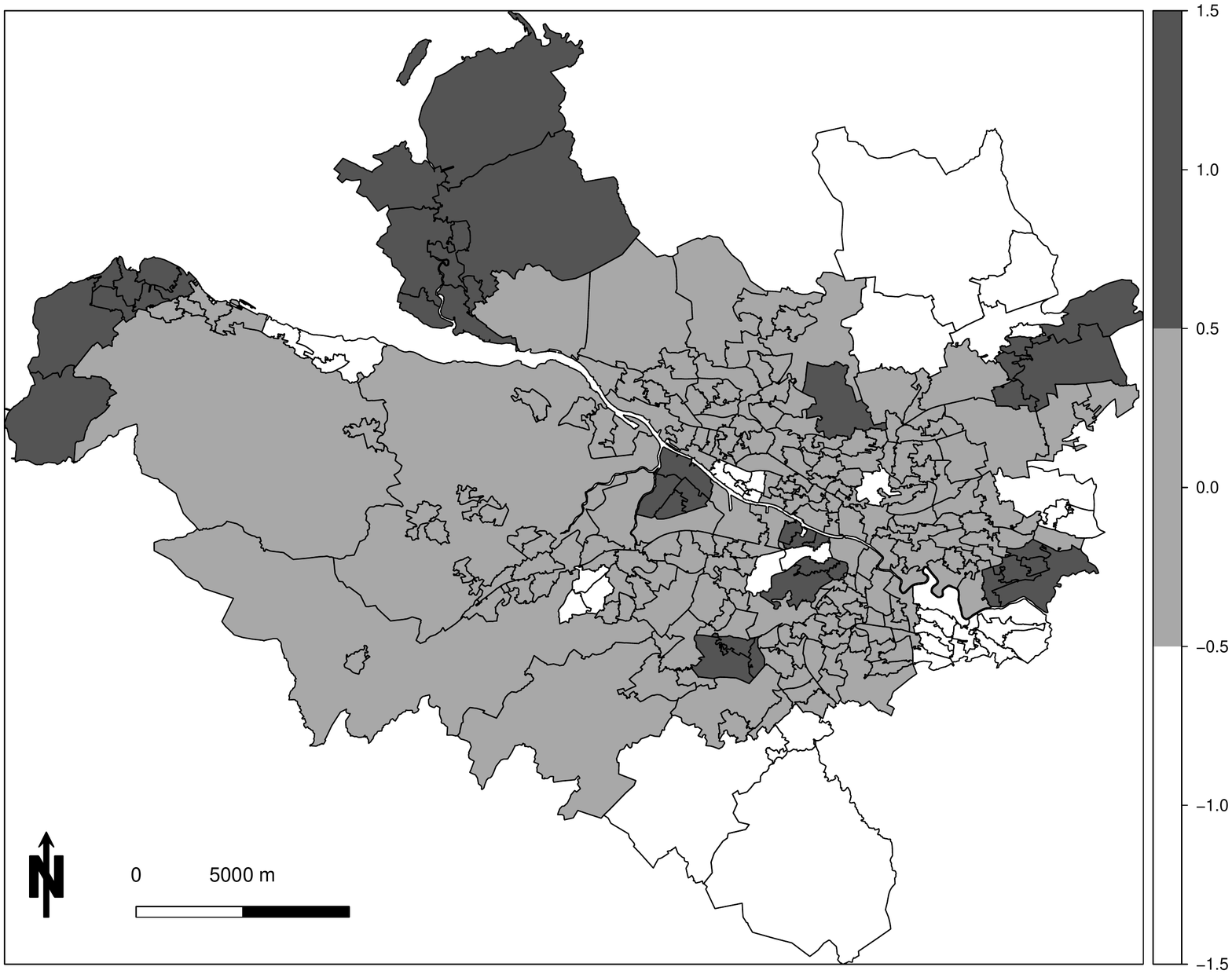}
\caption{Plot of the simulated cluster structure in the Greater Glasgow study region.}
\label{fig:SimClust}
\end{center}
\end{figure}

Clustered disease data were generated according to the template shown in Figure \ref{fig:SimClust}.  The template consists of 19 clusters of different sizes, which include the large cluster shaded in light grey and the 18 smaller clusters shaded in either white or dark grey, some of which are singletons.  Disease data were generated under this template from model (\ref{eq:PoissonLL}), with the simplification that no covariates were included.  The random effects were generated from a multivariate Gaussian distribution with a spatially correlated precision matrix, which was defined by the CAR model proposed by \cite{leroux99}. Note, the intrinsic model (\ref{eq:ICAR}) is not used for the data generation because its precision matrix is singular. Clustered disease data were obtained by specifying a piecewise constant mean function for \bd{\phi}, which follows the template shown in Figure \ref{fig:SimClust}.  The values in Figure \ref{fig:SimClust} are multiplied by $C$, where larger values of $C$ represent larger differences between the clusters, which should thus be easier to identify.  Values of $C = 0, 0.5,1$ are used in this study, where $C=0$ corresponds to a spatially smooth risk surface where one would hope to identify a single cluster covering the entire study region. For the analyses described in this section the expected disease counts are set equal to those from the respiratory disease motivating application. However, a sensitivity analysis assessing the robustness of our methodology to changing $\mathbf{E}$ is presented in Section 4 of the supplementary material available at Biostatistics online.\\

Each simulated data set consists of the study data and three sets of prior data, with the latter being used for the prior elicitation step. To allow for the fact that the log risk surfaces for the study and prior data sets are unlikely to be identical, uniform random noise was added to the random effects from the three prior data sets, which corresponds to multiplicative random noise on the risk scale. Greater levels of noise were added to the prior data the further away in time it was from the study data, and the levels of random noise for the three prior data sets were  U$[-0.1, 0,1]$, U$[-0.15, 0.15]$ and U$[-0.2, 0.2]$, and were chosen to match the correlations between the study and prior data sets in the motivating application in Section 5. Five hundred datasets were generated for each of the three scenarios ($C=0,0.5, 1$), and the model proposed here was compared against the  Besag-York-Molli\'{e} (BYM, \cite{besag91}) model, which is commonly used in disease mapping. To identify clusters in the fitted risk surface the posterior classification approach described in \cite{charras12} and \cite{charras13} was implemented, which is based on a Bayesian regularisation for Gaussian mixtures (\cite{fraley2007}). However, this approach does not produce spatially contiguous clusters, so a further post-processing step was implemented to partition the clusters identified into spatially contiguous groups. We note that we have not compared our approach to a method such as \cite{knorr00} or \cite{charras12}, because software to implement these complex estimation methods is not publicly available, and also because they use different inferential frameworks which may affect the results. In contrast, the BYM model was implemented using INLA, which is the inferential approach adopted here.

\subsection{Results}
\label{sec:ModelResults}
\begin{figure}[h!]
\begin{center}
\includegraphics[scale=0.65]{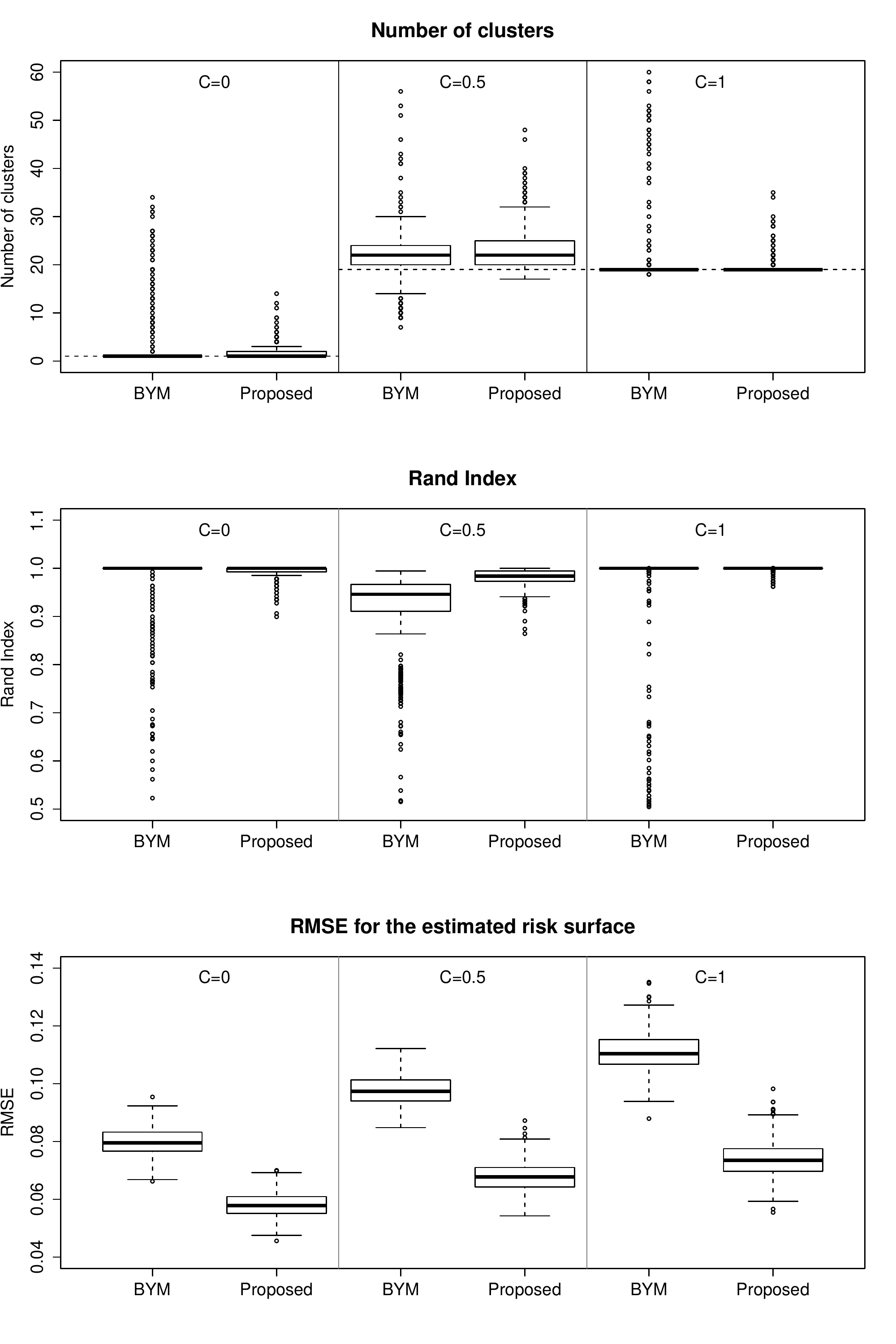}
\caption{Summary of the simulation study results. The top, middle and bottom panels display boxplots of the estimated  number of clusters, the Rand Index and the root mean square error of the estimated risk surface for the BYM model and the model proposed here. The results relate to $C=0$ (left), $C=0.5$ (middle) and $C=1$ (right). In the top panel the dashed lines represent the true number of clusters.}
\label{fig:sim}
\end{center}
\end{figure}

The results of the study are summarised in Figure \ref{fig:sim}, which displays a comparison of the relative performances of the approach proposed here and the BYM model using three different metrics. The accuracy of the risk surfaces estimated by both approaches is quantified by their root mean square error (RMSE), while the correctness of the estimated cluster structures is quantified by both the number of clusters identified and the Rand Index (\cite{rand71}) between the true and estimated cluster structures. The latter is a measure of the similarity between two cluster structures and lies in the interval $[0,1]$. It is computed as the proportion of pairs of areal units classified either in the same or in different clusters by both methods, that is the proportion of pairwise agreements between the two methods. A value of one indicates complete agreement between the two cluster configurations, while a value of zero indicates that no pair of areal units are classified in the same way under both configurations.\\

The top panel of Figure \ref{fig:sim} shows boxplots of the numbers of clusters estimated by each method in the 500 simulated data sets, where the true values of 1 (when $C=0$) and 19 (when $C=0.5,1$) are represented by dashed lines. The middle panel displays boxplots of the Rand index for all simulated data sets, while the bottom panel shows the RMSE values for the estimated risk surface. The top panel shows that when $C=0$ or $C=1$ both methods estimate the correct number of clusters on average, while when $C=0.5$ the median values are slightly high at 22 for both models. The median Rand Index values are equal to one for both models when $C=0$ or $C=1$, while when $C=0.5$ the values are 0.9461   and 0.9840 for the BYM model and that proposed here. However, there are a small number of data sets for which the BYM model produces poor results as measured by the above two metrics, while this phenomenon is much less pronounced for the model proposed here. Finally, The figure shows that the RMSE is always lower using the method proposed here compared with the BYM model, with reductions of 27.3$\%$ ($C=0$), 30.5$\%$ ($C=0.5$), and 33.3$\%$ ($C=1$) respectively.

\section{Motivating application}
\label{sec:Application}
\subsection{Study design}
\label{sec:Data}
The study region is the Greater Glasgow and Clyde health board, which contains the city of Glasgow in the east and the river Clyde estuary in the west.  Glasgow is the largest city in Scotland, with a population of around 600,000 people.  It is split into $n=271$ administrative units known as intermediate geographies (IGs), which contain populations of between 2,244 and 10877 people with a medan value of 4,239. The disease data are the numbers of hospital admissions with a primary diagnosis of respiratory disease in each IG in 2011, which corresponds to the International Classification of Disease tenth revision codes J00-J99 and R09.1. The expected numbers of hospital admissions  were calculated using external standardisation, based on age and sex adjusted rates for the whole of Scotland.    The top panel of Figure \ref{fig:BestClust} displays the Standardised Incidence Ratio (SIR) for respiratory hospital admission, which is the ratio of the observed to the expected numbers of cases. The figure shows that there are regions of high risk in the east of the city and directly south of the river, which contain the heavily deprived neighbourhoods of Easterhouse and Govan.  In contrast, areas in the centre (just north of the river) and far south of the study region exhibit much lower risks, which are the affluent West End and Giffnock districts of the city.

\subsection{Results}
\label{sec:AppResults}
\begin{figure}[h!]
\begin{center}
\includegraphics[scale=0.5]{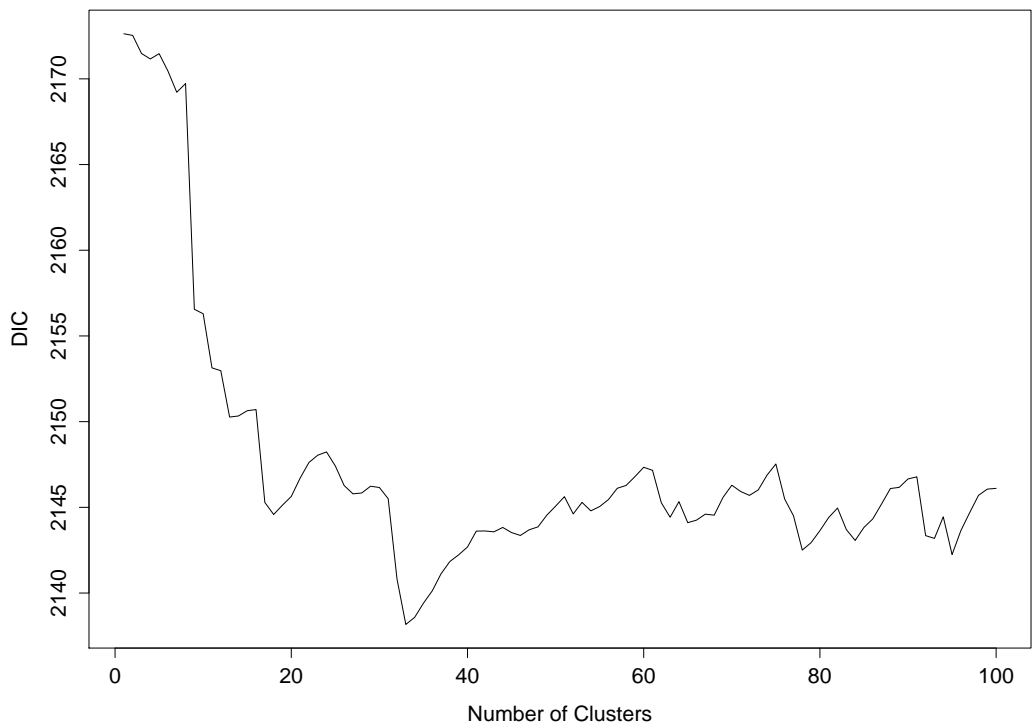}
\caption{Plot of the Deviance Information Criterion for models with between 1 and 100 clusters.}
\label{fig:DICPlot}
\end{center}
\end{figure}

The two-stage clustering model proposed in Section \ref{sec:Method} was applied to these data, where the prior elicitation step was based on respiratory disease data from 2008 to 2010. The fitted risk surfaces for these data sets exhibit similar spatial patterns to the 2011 study data, with Pearson's correlation coefficients of 0.86 (2010 data), 0.84 (2009 data) and 0.82 (2008 data) respectively. Model (\ref{eq:ClustModel}) was applied to the data with between 1 and 100 clusters, and Figure \ref{fig:DICPlot} shows the DIC values for these models. The model with 33 clusters minimises the DIC, while only models with between 32 and 38 clusters are within 4 of this minimum DIC value.\\

\begin{figure}[!h]
\begin{center}
\includegraphics[scale=0.45]{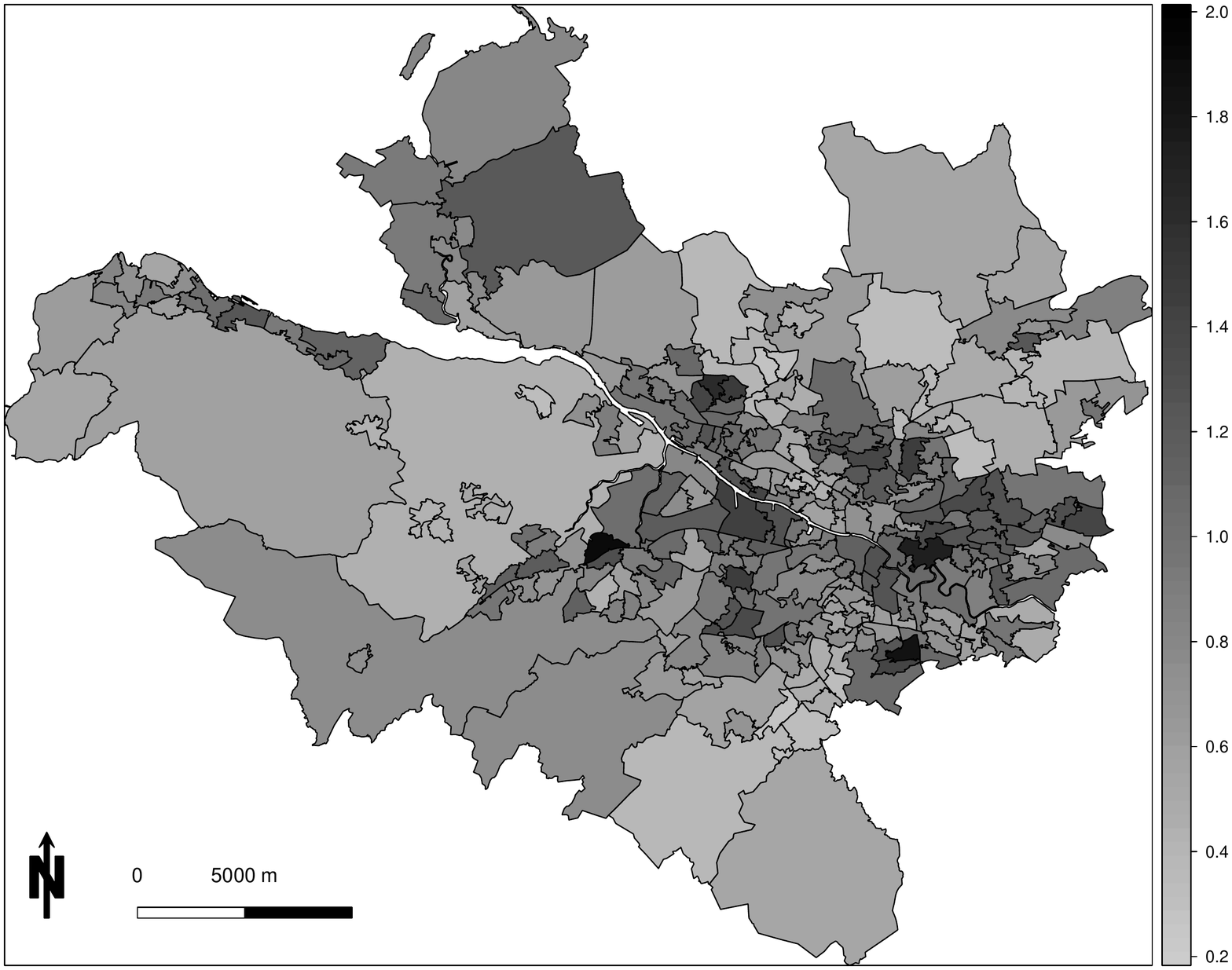}
\includegraphics[scale=0.45]{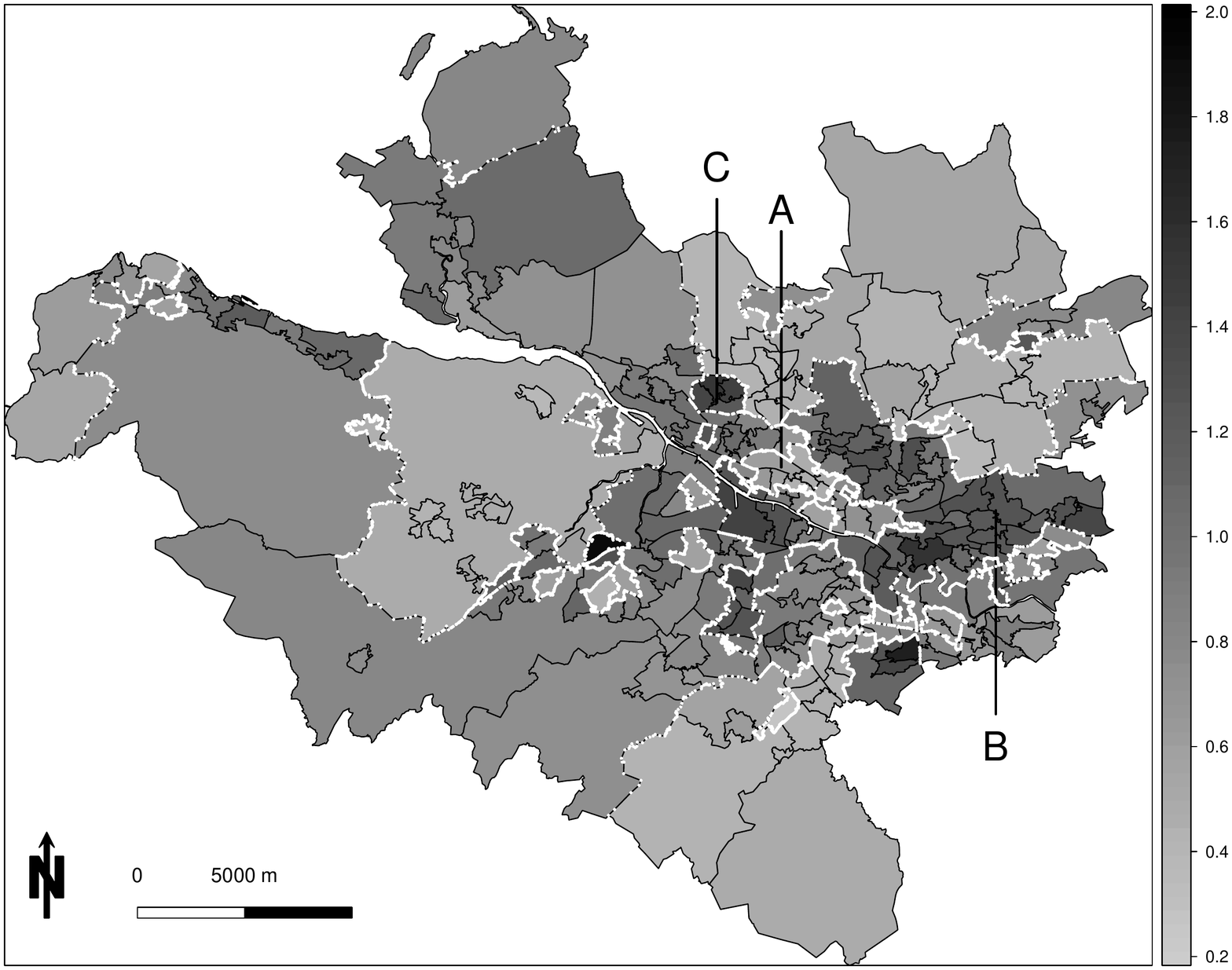}
\caption{The top panel displays the standardised incidence ratio (grey-scale) for respiratory disease hospitalisation in 2011 in Greater Glasgow. The bottom plot displays the estimated risk surface (grey-scale) from the model with 33 clusters (white dots). The labels A, B and C represent prominent clusters that have been identified.}\label{figure data1}
\label{fig:BestClust}
\end{center}
\end{figure}

The estimated risk surface (grey-scale) and cluster structure (white dots) are displayed in the bottom panel of Figure \ref{fig:BestClust},  which has the same scale as the SIR plot in the top panel of that figure. The majority of the clusters identified appear to exhibit different  risks compared with neighbouring areal units, and three of the prominent features of the risk map are highlighted A, B and C. Cluster A is the low-risk `west end' of Glasgow, which is one of the most affluent parts of the city. The large high-risk cluster denoted by B contains a number of the most deprived neighbourhoods of Glasgow, including Easterhouse in the east and Springburn and Summerston in the North. Finally, cluster A is the deprived suburb of Drumchapel, which exhibits elevated risks compared with the affluent Bearsden area to the north east. The main driver of these cluster configurations is socio-economic deprivation, which is well known to have a large effect on population health. The high-risk areas in Figure \ref{fig:BestClust} typically exhibit high levels of socio-economic deprivation, where as low-risk areas are more affluent. One could of course include a covariate measuring deprivation in the regression model to account for this, but it would explain the spatial pattern in respiratory disease risk rather than identifying the spatial extent of the high-risk clusters.

\section{Conclusion}
\label{sec:Conclusion}

The aim of this paper was to identify discontinuities in the spatial pattern of disease risk, which corresponds to the identification of clusters exhibiting both elevated and reduced risks.  The methodology we have developed is a fusion of spatially-adapted hierarchical agglomerative clustering techniques with conditional autoregressive models, the former being applied to  data quantifying disease risk prior to the study period to elicit a set of candidate cluster structures for the study data. Separate spatial random effects models are then applied to the study data for each candidate cluster structure, and the choice of the best cluster structure is treated as a model comparison problem. The Bayesian hierarchical models fitted in the second stage represent disease risk with a  linear combination of a spatially smooth intrinsic CAR model and a piecewise constant cluster model, which allows disease risk to evolve smoothly within a cluster with a disjoint multiplicative jump between clusters. Removing the CAR component of the model would assume a constant disease risk within a cluster, which is unlikely to be true in general.\\

The model comparison approach adopted here does not estimate the cluster structure simultaneously with the risk surface as is done by \cite{knorr00}, which ignores the uncertainty about the number of clusters in the estimation procedure. However, this two-stage approach is easy to implement and makes the identification of the `final' cluster structure straightforward, which is not always the case for approaches such as \cite{knorr00} which may produce a different cluster structure for each MCMC iteration.\\

The simulation study presented in Section four and the supplementary material showed that our model generally performs well, in particular outperforming the BYM model with a posterior classification step using a Bayesian regularisation for Gaussian mixtures. Improved performance was observed for both risk estimation and cluster identification, which is most likely to be because our approach attempts to estimate the cluster structure in the data. In contrast,  the posterior classification approach estimates  a smooth risk surface using the BYM model, and attempts to identify clusters from that smoothed surface. The studies we have conducted also suggest that our method performs well for diseases with moderate to large numbers of cases, but that when the number of cases in each areal unitis less than 25 it, like other methods, is likely to be less accurate at identifying the correct cluster structure.\\

There is scope to extend this method in two main ways. The approach proposed here models spatial discontinuities (clusters) in risk via the mean function using a piecewise constant fixed effect, which contrasts with the majority of the open boundary literature which achieves this by modelling the correlation structure in the random effects (see \cite{lu07} and \cite{leemitch12}).  Therefore adapting Stage 2 of the approach proposed here to identify clusters via the correlation structure of the random effects is a natural extension, and would provide an interesting comparison with what we have proposed here. The second avenue for future work is to extend these methods into the spatio-temporal domain, thus allowing policy makers to identify whether a health intervention has had an effect in reducing disease risk in a high risk cluster.

\section*{Supplementary material}
The supplementary material  available at Biostatistics online contains a number of additional simulation results, as well as functionality to implement the two-stage model proposed here.

\section*{Acknowledgements}
The authors gratefully acknowledge helpful suggestions made by the associate editor and two referees, which have improved the motivation for and content of this paper. The work of the first author was funded by the Carnegie Trust.

\bibliographystyle{chicago}
\bibliography{LitRev}
\nocite{*}

\end{document}